\documentclass[fleqn]{article}
\usepackage{espcrc2}
\usepackage{psfig}

\title{Quantum Interference Phenomena Between Impurity States in
d-wave Superconductors}
\author{U. Michelucci\address[EKM]{Theoretische Physik III,
Elektronische Korrelationen und 
Magnetismus, Institut f\"ur Physik, Universit\"at Augsburg, 86135 
Augsburg, Germany}, F. Venturini\address{Walther Meissner
Institut, Bayerische Akademie der 
Wissenschaften, 85748 Garching, Germany}, and A.P. Kampf\addressmark[EKM]}

\begin{document}

\begin{abstract}
We investigate the mutual influence of impurities in two-dimensional
$d$-wave 
superconductors involving self-consistent solutions of the
Bogoliubov-de Gennes equations.
The local order parameter suppression, the local
density of states (LDOS) as well as the interference of
impurity-induced structures are analyzed. 
We employ an impurity position averaging scheme for the DOS that
does not neglect these interference effects, as the commonly used 
$T$-matrix approaches do.
\end{abstract}

\maketitle

In recent years scanning tunneling microscopy (STM) has become an
excellent high resolution probe of the electronic structure near
defects. In particular it has been used
to investigate the effects of individual Zn and Ni atoms in the
$d$-wave superconductor
Bi$_2$Sr$_2$CaCu$_2$O$_{8+\delta}$ (BSCCO) \cite{Pan}.
These experiments show the appearance of a quasiparticle bound state
with a fourfold symmetry around each impurity, reflecting directly
the anisotropic 
$d_{x^2-y^2}$ pairing symmetry \cite{Pan}.

The existence of impurity-induced bound states was 
predicted early on by Balatsky {\it et al.}  using the
$T$-matrix approximation 
\cite{Balatsky,Salkola1}. 
Often, local variations in the local order parameter (LOP) dynamically
generated by the impurity potential 
are neglected. Since the LOP varies only on a
coherence length scale, which is very short ($\xi \approx 20$
\AA) in high-$T_c$ superconductors, the overall effect of this
variation was argued to be negligible.  
Recent calculations, however, indicate
otherwise \cite{Atkinson,Hirschfeld} showing that the inclusion of the
order parameter relaxation allows to achieve a better agreement with
microwave conductivity data.

In pure $d_{x^2-y^2}$ superconductors the averaged density of states
(DOS) $\rho(E)$ vanishes linearly for $|E|  
\rightarrow 0$. With impurities the self consistent T-matrix  
approximation \cite{schmitt} leads to a finite DOS at $E=0$, non
perturbative approaches have variously predicted that $\rho(E)$
vanishes \cite{ziegler2} or that it diverges \cite{pepin}.
These issues are still
controversial and some efforts are needed to settle the low
energy DOS issue in the presence of a finite concentration of
impurities. 

In this paper we include the local gap suppression of the order
parameter and examine the LDOS for two nearby non-magnetic impurities
and for a finite
impurity concentration to explore how impurity induced structures can
interfere with each other, and to what extent this influences the
impurity position averaged DOS.

We start from a pairing Hamiltonian for
electrons hopping on a square lattice with nearest
neighbor hopping matrix element $t$, and a superconducting bond order
parameter $\Delta_{i,j}$:
\begin{equation}
\begin{array}{lll}
\medskip {\cal H} & = & -t \displaystyle \sum_{\langle i,j
\rangle,\sigma} c_{i,\sigma}^\dag 
c_{j,\sigma} +\displaystyle  \sum_{\{ l \},\sigma} U_l
c_{l,\sigma}^\dag  
c_{l,\sigma} \\
&&+\displaystyle \sum_{\langle i,j \rangle} \{
\Delta_{i,j} c_{i,\uparrow}^\dag c_{j,\downarrow}^\dag + h.c.
\} - \mu \displaystyle \sum_{i,\sigma} c_{i,\sigma}^\dag 
c_{i,\sigma}.
\end{array}
\label{ham}
\end{equation}
The non-magnetic impurity potential is summed over impurity positions
($\{l\}$) and
$U_l$ is the local potential at the impurity site $l$ . Energies will be 
measured in units of $t$ and lengths in units of the lattice constant $a$.
In what follows we set $\mu =0$ corresponding to half-filling since
the variation of the filling does not change our
conclusions. Note however that, even for $\mu=0$, ${\cal H}$ is not
particle-hole (PH) symmetric due to the presence of $U_l$.

We first consider  a single
impurity located at site $\hat {\i}$, and we choose $U_{\hat {\i}} = U_0$.
We perform the unitary transformation \cite{deGennes} 
\begin{equation}
\medskip c_{i,\sigma} = \displaystyle \sum_n \left(
\gamma_{n,\sigma} u_{i,n}+(-1)^\sigma \gamma_{n,-\sigma}^\dag v_{i,n}^{*}
\right)
\label{transf}
\end{equation}
and by imposing the
condition that the transformation (\ref{transf}) diagonalizes the
Hamiltonian (\ref{ham}) we obtain the Bogoliubov-de Gennes (BdG)
equations \cite{deGennes}
\begin{equation}
\displaystyle \sum_{j} \left(
        \begin{array}{ll}
                H_{i,j} & \Delta_{i,j} \\
                \Delta_{i,j}^* & H_{i,j}^* \\
        \end{array}
\right)
\left(
        \begin{array}{l}
        u_j^n \\
        v_j^n \\
        \end{array}
\right) =
E_n
\left(
        \begin{array}{l}
        u_i^n \\
        v_i^n \\
        \end{array}
\right).
\label{bdgeq}
\end{equation}
Here $\Delta_{i,j}$ is the self-consistent mean field solution for the
order parameter 
\begin{equation}
\Delta_{i,j} = V_{i,j} \langle c_{i,\uparrow}
c_{j,\downarrow} \rangle. 
\label{selfcons}
\end{equation}
and $V_{i,j}$ is the attractive pairing interaction. 

\begin{figure}
\psfig{file=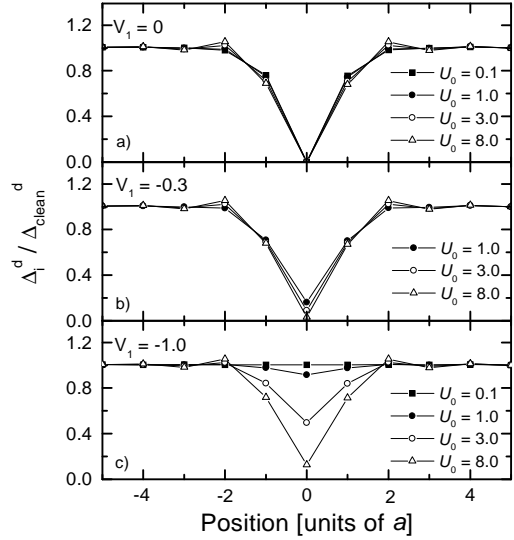,width=7cm}
\caption{Normalized $d$-wave component of the order parameter
along the (1,0) direction for different impurity potential strengths 
$U_0$. The pairing interaction is set to $V_0=-1$. 
Panels a,b,c correspond to different values of $V_1$ to account for
the suppression of the pairing interaction near the impurity.
The position is measured with respect to the impurity site.} 
\label{Figure_gap}
\end{figure}

We solve numerically the self-consistent BdG equations (\ref{bdgeq})
and (\ref{selfcons}) on lattices of 20$\times$20 sites and with
open boundary conditions. 
For the pairing interaction we choose
\begin{equation}
V_{i,j} = V_0 \hat \delta_{i,j} +V_1 (1-\hat \delta_{i,j})
\end{equation}
where $\hat \delta_{i,j}=1$ when $i,j \neq \hat {\i}$ and $i,j$ nearest
neighbors. This form is motivated by the assumption 
that the pairing interaction is reduced around the
impurity, and we therefore consider parameters with $|V_1| <
|V_0|$. Note that for $V_1=V_0<0$ this form of the 
pairing leads to a $d_{x^2-y^2}$ structure of the order parameter.
The local $d$-wave component of $\Delta_{i,j}$ is obtained from:
\begin{equation}
\medskip \Delta^d_i = \displaystyle \frac{1}{2} \left(
\Delta_{i,i+x}+\Delta_{i,i-x}-\Delta_{i,i+y}-\Delta_{i,i-y}
\right).
\label{harmonics}
\end{equation}
In the clean case $\Delta^d_{i}$ is uniform in space and has a value
of $\Delta^d_{i}\approx 0.48$ for $V_0=-1$. 
Modifications due to impurities or due
to the border extend in space on a length scale of the order of the
coherence length $\xi$. For our set of parameters we estimate
$\xi \approx 4$ using $\xi=v_F/\Delta_0$, where $v_F$ is the Fermi velocity and
$\Delta_0$ is the maximum of LOP. The value of $V_0$ and
consequently the value of $\Delta_i^d$ are chosen to render $\xi$
small enough on a 20$\times$20 system so that a sizeable portion
of the lattice can still be considered as representative of the bulk
of the system.
Fig. \ref{Figure_gap} shows $\Delta_i^d$ near the impurity site along the
(1,0) direction for various values of $V_1$ and $U_0$. 
A common feature is a strong suppression of the
$d$-wave component of the order parameter around the impurity site
on a range of roughly 3 lattice
constants, in agreement with the estimated coherence
length $\xi$. 
The LOP along the (1,1) direction is modified
on the same length scale.
Note that in the case of $V_1=0$ we have $\Delta_{\hat {\i}}^d=0$
since $\Delta_{\hat {\i},\hat {\i}\pm x} = \Delta_{\hat {\i},\hat {\i}\pm
y}=0$. 

\begin{figure}
\centerline{\psfig{file=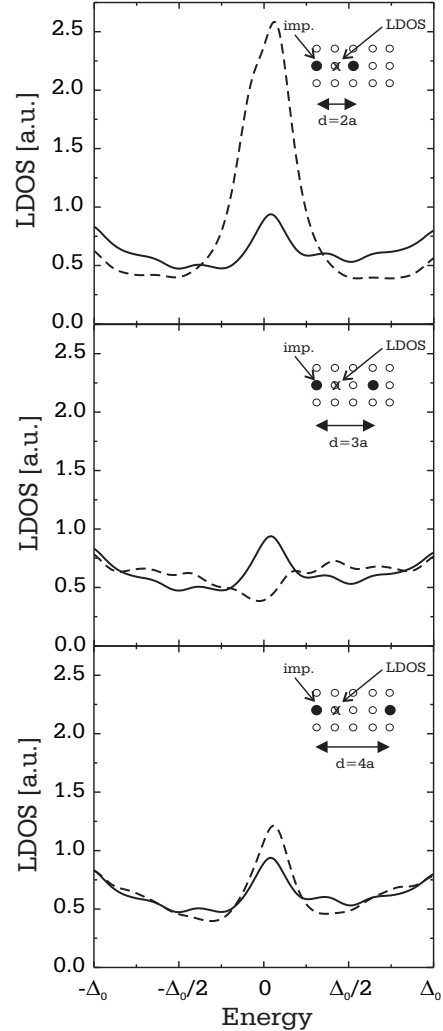,width=6cm}}
\medskip
\caption{Local density of states in the presence of two impurities
measured near one impurity site. The 
continuous lines are the calculated densities of states with only one
impurity, the dashed ones when the second impurity is added. The LDOS
is always measured immediately on the adjacent site of the first
impurity. The parameters used are: $\Gamma=0.1$, $V_0=-1$, $V_1=0$,
$U_0=8$.}  
\label{Figure_LDOS}
\end{figure}

\begin{figure}[hbt]
\centerline{\psfig{file=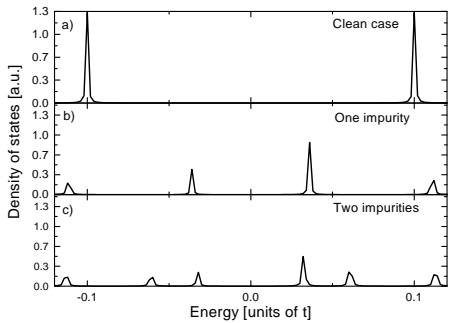,width=7cm,silent=}}
\caption{Plot of the DOS with $\Gamma=0.001$ in the region
$|E|<0.1$. a) Clean case result; b) with one impurity; c) with
two impurities located three sites away one from each other in the
(1,0) direction, for $U_0 = 8$, $V_0=-1$, $V_1=0$.}
\label{Figure_En}
\end{figure}

The LDOS as obtained from 
\begin{equation}
\rho_i(E) = \sum_n \left[
|u_i^n|^2 \delta(E-E_n) + |v_i^n|^2 \delta(E+E_n) 
\right]
\label{LDOS}
\end{equation}
near the impurity is plotted in
Fig. \ref{Figure_LDOS} as the continuous line. 
In Eq. (\ref{LDOS}) the sum is over the eigenstates $n$ of the Hamiltonian
(\ref{ham}) and $E_n$ denote the corresponding eigenenergies. In the
numerical calculation the delta functions present 
in Eq. (\ref{LDOS}) are approximated by Lorentzians
with a width $\Gamma$. Typically we use $\Gamma/t =
0.1$. 
On the impurity site the LDOS is strongly reduced while on the nearest
neighbor sites we find two peaks around zero energy in the LDOS,
associated with two resonant quasiparticle bound states with energies
$\pm \omega_R$,  (see
Fig. \ref{Figure_En} where the DOS is calculated with
$\Gamma=0.001$). The two peaks
are not resolved in Fig. \ref{Figure_LDOS} due to the large broadening
$\Gamma=0.1$, and only one peak is observed around zero energy.  
This peak is, also if not evident from the figure,
asymmetric revealing the underlying double peak structure at $\pm
\omega_R$. 
Measuring the distance variation of the LDOS around the impurity at
the resonance energy reveals also the well known fourfold symmetric
structure characteristic for a $d_{x^2-y^2}$ superconductor \cite{Pan}. 

In the next step we add a second impurity and explore the mutual
influence of the 
two impurities. The resulting LDOS at a selected site between the two
impurities is 
shown in Fig. \ref{Figure_LDOS} (dashed line), for different impurity
distances along the (1,0) direction.
With one impurity we observe a resonance peak at very low energy
$\omega_R$ 
corresponding to the lowest positive eigenvalues of the system that has an
energy of $E_0 = \omega_R \approx 0.04$. When a
second impurity is added at a distance $d=2$ from the first, the peak
is strongly enhanced by a factor of roughly 2.5. Its asymmetry is
again reflecting the underlying resonant impurity states at positive
and negative energies (see Fig. \ref{Figure_En}). When the second
impurity is moved one site further away, the peak
disappears leaving a dip around zero energy. The peak is
again recovered when the distance is increased further ($d=4$).
This alternating variation is naturally explained as an interference
phenomenon due to the overlap of the bound state wave functions near
each impurity.
In order to demonstrate in detail what is happening at low
energies when one or more impurities are added to the system we plot
in Fig. \ref{Figure_En}  the DOS
for $\Gamma=10^{-3}$. In Fig. \ref{Figure_En}.a we show the 
clean case as a reference; the spikes at $|E| \approx 0.1$ are the
lowest energy eigenvalues for a 20x20 system with open boundary
conditions; note the symmetry of the LDOS in the clean case: for $U_0=0$
the Hamiltonian is particle-hole symmetric for $\mu=0$. 
When we add an impurity at the center of 
the system we observe the appearance of two additional eigenvalues
with asymmetric weights at
energies $|E|\approx 0.04$. The asymmetry results from the broken
particle-hole symmetry due to the presence of the impurity.
When we add a second impurity near the first one
(see Fig. \ref{Figure_En}.c) the 
number of eigenvalues in the region $|E|<0.1$ is doubled.
Thus, by adding impurities one by one an impurity band of finite width
is formed due to the overlaps of individual impurity induced
quasiparticle bound states \cite{Joyint}.

When a self-consistent $T$-matrix approach is used to calculate the
impurity position averaged DOS the interaction effects between impurities are
neglected due to the selection of non-crossing diagrams for impurity
scattering in the standard $T$-matrix approximation. Therefore, the
mutual influence of impurities as observed above is missed in the
conventional $T$-matrix approach.
\begin{equation}
\rho_{av} (E) = \frac{1}{N_c} \sum_m \rho^{(m)} (E)
\end{equation}

\begin{figure}
\centerline{\psfig{file=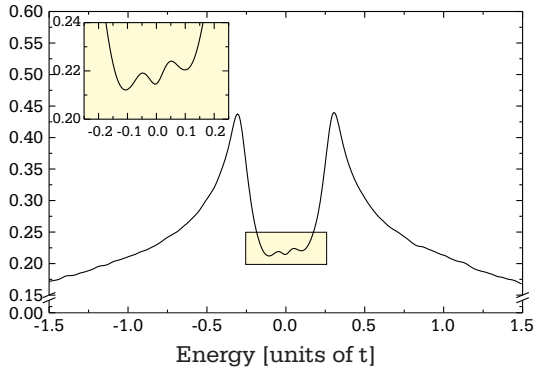,width=7cm}}
\caption{Averaged DOS for an impurity
concentration of $n_i=20 \%$. The inset is a zoom of
the low energy part of the calculated DOS.}
\label{Figure_average}
\end{figure}
An obvious question to ask is to what extent the interference effects
and the impurity band formation may survive an appropriate impurity
position averaging scheme.
In order to answer this question we introduce in the
system a fixed number of impurities in random positions and
calculate the LDOS in the center of the lattice
$\rho^{(1)}(E)$. 
We keep the site for the measurement of the LDOS fixed for all
impurity configurations with the same impurity concentration.
The averaged DOS is calculated according to:
where $m$ enumerates an impurity configuration and $N_c$ is the total
number of impurity configurations. Typically we choose $N_c \sim
10^2$, convergence is usually achieved already for $N_c>50$.
In Fig. \ref{Figure_average} $\rho_{av} (E)$ is shown for a large
impurity concentration $n_i = 20 \%$ to highlight a new low energy
structure in the averaged DOS that is not appearing in the usual
$T$-matrix calculation, that leads to a flat and finite DOS
around zero energy. 
The new feature that we find consists of an asymmetric
structure centered around zero energy.
This previously unresolved feature in the DOS survives also when $n_i$
decreases to a few percent although it becomes less pronounced; its origin
is clearly related
to the overlap of the impurities induced bound states and the
accompanying impurity band formation.
The above discussed evolution of the impurity band naturally suggests
furthermore that the DOS at zero energy will remain zero even for a
finite impurity concentration.

%
%

In conclusion we have outlined an extension of the BdG formalism to
two dimensional 
superconductors with a finite concentration of impurities. We have
shown how impurities interact between each other giving rise to
interference phenomena and leading eventually to an impurity band
formation due to the finite overlap of quasiparticle bound
states. These phenomena survive for an appropriately chosen 
impurity position averaging scheme leading to a double peak
structure in the averaged DOS at low energies which may be resolvable
in STM measurements.  

This work was partially supported by the Deutsche
Forschungsgemeinschaft through SFB 484.

%
%

\end{document}